\documentclass[sigconf,nonacm]{acmart}
\renewcommand\footnotetextcopyrightpermission[1]{}

\usepackage{enumitem}
\usepackage{comment}
\usepackage{listings}
\usepackage{tcolorbox}
\tcbuselibrary{listings}
\usepackage{xcolor}

\usepackage{algpseudocode}
\usepackage{amsmath}
\usepackage{booktabs} % 用于更好的表格线条
\usepackage{multirow}
\usepackage{csquotes}
\usepackage{tabularx}
\usepackage{booktabs}
\usepackage{ltablex}
\usepackage{array}
\usepackage{ifsym}
\usepackage{amsmath}
\usepackage{xcolor}
\usepackage{graphicx}
\usepackage{float}
\usepackage{subcaption}
\usepackage[linesnumbered,ruled,vlined]{algorithm2e} 
\SetKwInput{KwInput}{Input}
\SetKwInput{KwOutput}{Output}
\SetKwInput{KwHyperparameters}{Hyperparameters}
\SetKwInput{KwInitialize}{Initialize}
\newlength{\lstoffset}
\usepackage{xcolor}
\usepackage{listings}
\usepackage{float}

\usepackage{dblfloatfix}

\lstdefinestyle{prompt-verilog}{
  language=Verilog,
  basicstyle=\ttfamily\scriptsize,
  keywordstyle=\bfseries\color{blue},
  backgroundcolor=\color{gray!5},
  frame=single,
  rulecolor=\color{black!30},
  framesep=3pt,
  columns=fullflexible,
  keepspaces=true
}

\lstdefinestyle{prompt-scala}{
  language=Scala,
  basicstyle=\ttfamily\scriptsize,
  backgroundcolor=\color{gray!5},
  frame=single,
  rulecolor=\color{black!30},
  framesep=3pt,
  columns=fullflexible,
  keepspaces=true
}

\lstdefinestyle{prompt-asm}{
  language={[x86masm]Assembler},
  basicstyle=\ttfamily\scriptsize,
  backgroundcolor=\color{gray!5},
  frame=single,
  rulecolor=\color{black!30},
  framesep=3pt,
  columns=fullflexible,
  keepspaces=true
}

\newcolumntype{C}[1]{>{\centering\arraybackslash}p{#1}}
\AtBeginDocument{%
  }

\begin{document}

\settopmatter{printacmref=false}

\acmConference[ICCAD '26]{IEEE/ACM International Conference on Computer-Aided Design}{November 8--12, 2026}{San Jose, California, USA}
\acmBooktitle{Proceedings of the 2026 IEEE/ACM International Conference on Computer-Aided Design (ICCAD '26), November 8--12, 2026, San Jose, California, USA}
\acmYear{2026}
\copyrightyear{2026}
\title{When Fuzzing Meets Understanding: LLM-Driven Semantic Test Generation for RTL Verification}

\author{Kun Wang}
\email{wangkun22@mails.ucas.ac.cn}
\affiliation{%
  \institution{Institute of Computing Technology, Chinese Academy of Sciences}
  \country{China}
}

\author{Cangyuan Li}
\email{licangyuan@ict.ac.cn}
\affiliation{%
  \institution{Institute of Computing Technology, Chinese Academy of Sciences}
  \country{China}
}

\author{Kaiyan Chang}
\email{changkaiyan@live.com}
\affiliation{%
  \institution{Institute of Computing Technology, Chinese Academy of Sciences}
  \country{China}
}

\author{Siyang Cai}
\email{caisiyang23@mails.ucas.ac.cn}
\affiliation{%
  \institution{School of Advanced Interdisciplinary Sciences, University of Chinese Academy of Sciences}
  \country{China}
}

\author{Yinhe Han}
\email{yinhes@ict.ac.cn}
\affiliation{%
  \institution{Institute of Computing Technology, Chinese Academy of Sciences}
  \country{China}
}

\author{Ying Wang}
\email{wangying2009@ict.ac.cn}
\affiliation{%
  \institution{Institute of Computing Technology, Chinese Academy of Sciences}
  \country{China}
}

\renewcommand{\shortauthors}{Wang et al.}

\newcommand{\Equ}[1]{Equ.~\ref{#1}}
\maketitle

\begin{comment}
1. introduction 
    第一段 先说硬件验证在硬件开发中是非常重要的一环
    第二段 说硬件验证目前以为静态验证和基于模拟器的动态验证为主，其中覆盖率是衡量硬件验证效果的重要指标，基于覆盖率反馈的硬件模糊测试时重要的一环。
    第三段  然而现有的方法还存在一些主要的问题，在Rocket Core这些相对较大规模的CPU上覆盖率难以达到75%，
    第四段  LLM去做硬件验证的优势,或者说我们的insight

    Our main contributions are highlighted as follows:
    这部分需要参考LLM agent的文章去看看他们的agent是怎么描述的。
    1. 我们提出了RTLFuzzer 一个LLM agent框架用来做硬件模糊测试
    2. 实验结果显示RTLFuzzer能够达到什么效果
    3. 我们的代码开源

2. Background && Motivation 
    第一段讲 硬件模糊测试 
    第二段讲 大语言模型

    Motivation 部分就讲两个challenge  
3. Method

4. Evaluation

    4.1 Coverage metric 
        final coverage && 100K case coverage
    4.2 Bugs
    4.3 Ablation experiment
    
    随着现代处理器复杂度不断增高，给硬件验证带来了更多的挑战。最近基于覆盖率反馈的模糊测试已经成为了处理器验证中一种重要的手段。然而我们观察现在对大规模 CPU进行模糊测试时覆盖率是不够理想的，并且随着时间增加，测试效率会非常低下，为了提高模糊测试的效率，我们提出了一个基于大模型的agent框架，利用大模型对DUT系统相关知识的理解能力，针对性的生成测试用例。实验结果展示本身提出来的方法在覆盖率上达到了xx/%,和sota的相比在相同的xxK个case下，覆盖率能提高xx/%，证明了RTLFuzzer的有效性。
\end{comment}

\section{Abstract}

\begin{comment}

现代芯片日益增长的复杂性给硬件验证带来了严峻挑战。覆盖率引导的模糊测试已成为硬件验证的关键技术，然而现有先进模糊测试方法在复杂硬件设计上难以实现高覆盖率，且难以暴露深层边界案例错误。基于大语言模型（LLM）能够将测试生成导向特定代码区域的能力，我们提出了ChipFuzzer——一个基于智能体的高效硬件模糊测试框架。我们提出了一种结合CFG相似性索引与差分定位的模板提取方法，该方法显著提高了定向测试生成的成功率，从而有效提升代码覆盖率。此外，我们通过分析历史错误数据，提出三种定位易错代码区域的策略，通过将定向测试生成引导至这些区域，显著提升了错误发现能力。在包含三款知名开源CPU设计的ENCARSIA基准测试上的实验结果表明，ChipFuzzer实现了超过93%的代码覆盖率，并将错误检测率提升了21.1%。
\end{comment}
\begin{sloppypar}
The growing complexity of modern chips poses significant challenges to hardware verification. In recent years, coverage-guided fuzzing has emerged as a promising approach for improving verification efficiency. However,  existing hardware fuzzers still struggle to achieve high coverage and expose corner-case bugs, as they predominantly rely on heuristic strategies with limited ability to reason about the internal logic and semantic behavior of the design under test (DUT). In this work, we propose ChipFuzzer, a hardware fuzzing framework that leverages the semantic reasoning capabilities of large language models (LLMs) to improve fuzzing effectiveness. ChipFuzzer adopts a dual-stage workflow comprising  a Coverage-Guided stage and a Bug-Guided stage. In the Coverage-Guided stage, ChipFuzzer employs control-flow similarity and discrepancy analysis to guide LLM-driven testcase generation, thereby improving coverage.
 In the Bug-Guided stage, ChipFuzzer leverages historical bug data to identify bug-prone code regions and prioritize testcase generation for those regions, thus enhancing bug discovery efficiency. Experimental results on three open-source CPU designs show that ChipFuzzer improves average condition coverage by 5.8 percentage points and bug detection rate by 21.1 percentage points over the strongest baseline.

\end{sloppypar}


\section{Introduction}

As ICs continue to scale in complexity, verification has become the most resource-intensive and time-critical stage of the design cycle, often consuming more than 70\% of total project effort~\cite{farahmandi2020system}. Given the astronomical cost of post-silicon bug fixes, comprehensive pre-silicon verification is indispensable for ensuring design correctness and time-to-market success. To this end, researchers have developed numerous hardware 
verification techniques, which broadly fall into two 
categories: (1) formal verification, including theorem 
proving~\cite{cyrluk1994effective}, model checking~\cite{clarke2018handbook}, and 
information-flow tracking~\cite{hu2021hardware}; and (2) 
simulation-based verification, including random 
regression~\cite{naveh2007constraint} and hardware fuzzing~\cite{xu2024pathfuzz,trippel2022fuzzing, saravanan2024emergence,shen2025bmcfuzz}.
While formal methods promise exhaustive checking, they suffer from state space explosion in large-scale designs.  In contrast, coverage-guided fuzzing has gained traction as a scalable and effective methodology for hardware verification, where automatically generated testcases are iteratively refined to improve coverage and expose hidden corner-case bugs.

However, \textbf{hardware fuzzing today inherits fundamental limitations from its software origins}. Existing hardware fuzzers retain an input-centric methodology and often treat Hardware Description Language (HDL) signals and control paths as opaque variables, ignoring their rich structural semantics. This manifests as an excessive focus on test seed generation and mutation strategies, while neglecting the inherent structural characteristics of the DUT. Effective hardware verification, in contrast, should be grounded in a deep comprehension of the DUT's internal structure to generate targeted testcases. Although coverage-guided approaches have emerged as the predominant paradigm~\cite{xu2024pathfuzz} by tracking code coverage, their understanding of the DUT's functional semantics and internal architecture remains  superficial. As a result, they struggle to exercise hard-to-reach hardware behaviors, achieving only around 80\% coverage even on mid-scale CPUs such as RocketCore. Worse, as fuzzing progresses, the coverage growth curve  flattens~\cite{rostami2024beyond}, consuming massive compute resources for marginal gains. The generated testcases also lack diversity, being constrained by fixed seed pools and shallow mutation heuristics.

The emergence of LLMs offers a powerful new lens for rethinking this problem. LLMs trained on large corpora of HDL code and documentation have shown promise in reasoning about hardware intent, structure, and semantics. They can reason about signal dependencies, control flows, and corner conditions, which are  capabilities difficult for random mutation-based fuzzers to achieve. This motivates \textbf{semantic-aware fuzzing}, where test generation is context-driven rather than blind.

In this paper, we present ChipFuzzer, an LLM-driven hardware fuzzing framework  that rethinks hardware fuzzing through semantic-aware test generation. ChipFuzzer implements a \textbf{dual-stage verification methodology} to improve the efficiency of hardware fuzzing. In the \textbf{Coverage-Guided stage}, ChipFuzzer employs control-flow similarity indexing to retrieve relevant testcase templates and performs discrepancy analysis to 
iteratively guide LLM-driven testcase generation toward uncovered code regions. Beyond maximizing coverage, we further recognize that coverage alone does not guarantee robust verification. Many elusive design flaws lie in bug-prone patterns repeatedly seen across design iterations. Therefore, ChipFuzzer introduces the \textbf{Bug-Guided stage}. In this stage,  ChipFuzzer leverages 
historical bug data to identify bug-prone code regions and generate targeted testcases for these regions. ChipFuzzer 
further applies a semantic-aware seed fusion strategy that combines coverage-guided and bug-guided seeds to produce 
semantically richer testcases for exposing potential design flaws. This bug-guided stage shifts fuzzing beyond coverage-guided exploration alone toward more effective bug discovery.

Overall, our main contributions are as follows:
\begin{itemize}
    \item We propose ChipFuzzer, an LLM-driven hardware 
    fuzzing framework that leverages the semantic 
    reasoning capabilities of LLMs to improve fuzzing 
    effectiveness. ChipFuzzer implements a dual-stage 
    verification methodology consisting of a 
    Coverage-Guided stage and a Bug-Guided stage to 
    achieve both comprehensive coverage improvement 
    and effective bug discovery.
    \item For the Coverage-Guided stage, we introduce 
    control-flow similarity indexing for template 
    retrieval and discrepancy analysis for identifying 
    missing execution conditions, jointly guiding 
    LLM-driven testcase generation toward uncovered 
    code regions.
    \item For the Bug-Guided stage, we propose three 
    strategies for identifying bug-prone code regions 
    based on historical bug data, and a semantic-aware 
    seed fusion strategy that combines coverage-guided 
    and bug-guided seeds to improve bug detection 
    efficiency.
    \item Experimental results on three open-source CPU designs show that ChipFuzzer improves average condition coverage by 5.8 percentage points and bug detection rate by 21.1 percentage points over the strongest  baseline.
    \item We release our code for public access to 
    encourage further research: 
    \url{https://anonymous.4open.science/r/ChipFuzzer-212B}.
\end{itemize}
\section{Background \& Motivation}

\subsection{Hardware fuzzing}
\begin{figure}[htbp]
\includegraphics[width=\linewidth]{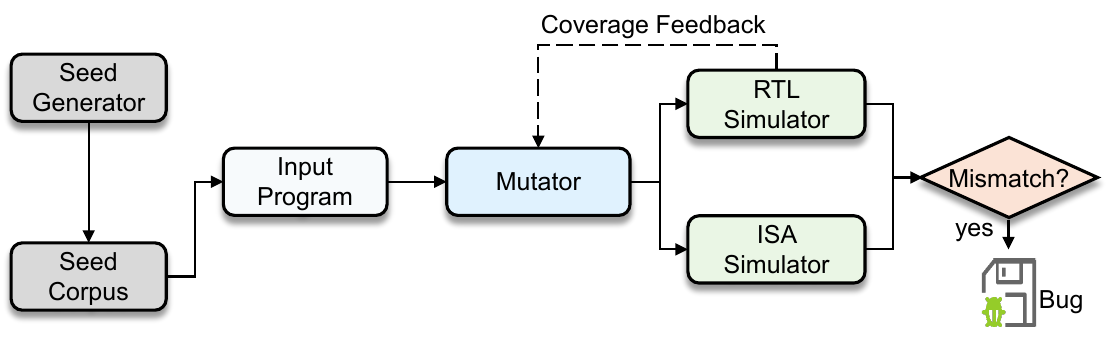} % Replace with the path to your image
    \caption{Typical hardware fuzzer flow.}
    \label{fig:fuzzer}
    
\end{figure}

Hardware fuzzing is an effective technique for exposing potential design flaws in the DUT, and its typical workflow is illustrated in Figure~\ref{fig:fuzzer}. Starting from an initial seed corpus, the fuzzer iteratively mutates testcases and executes them on the DUT to explore new behaviors. Generated testcases are evaluated based on coverage feedback, and high-value seeds are retained for further mutation. Through this coverage-guided feedback loop, hardware fuzzing can  incrementally expand behavioral exploration  and improve bug detection efficiency. Originating from software fuzzing, hardware fuzzing was introduced into RTL verification by early works such as RFUZZ~\cite{laeufer2018rfuzz}, which demonstrated that coverage-guided seed mutation can be effectively applied to hardware designs. Building on this direction, subsequent fuzzers such as DifuzzRTL~\cite{hur2021difuzzrtl}, MABFuzz~\cite{gohil2024mabfuzz}, and TheHuzz~\cite{kande2022thehuzz} further enhanced coverage efficiency by incorporating hardware-specific feedback signals. More recent methods, including HyPFuzz~\cite{chen2023hypfuzz}, Cascade~\cite{solt2024cascade}, BMCFuzz~\cite{shen2025bmcfuzz} and PSOFuzz~\cite{chen2023psofuzz}, integrate formal verification techniques or optimization methods to improve coverage.

\textbf{Limitations of Existing Hardware Fuzzers.} Despite significant advances in hardware fuzzing, existing approaches still rely predominantly on heuristic strategies without explicit causal reasoning about the DUT's internal logic and semantic behavior. Consequently, when a testcase succeeds or fails, the fuzzer cannot identify the conditions that activated or precluded the target behavior, nor can it systematically infer the input properties required to drive execution toward hard-to-reach code regions. By contrast, human verification engineers reason explicitly about the DUT's internal structure and triggering conditions when constructing targeted testcases. The absence of such causal reasoning leaves existing hardware fuzzers largely dependent on trial-and-error exploration, resulting in limited coverage and low verification efficiency in complex CPU designs.

\subsection{Large Language Models}
LLMs have achieved remarkable success in natural language processing (NLP). Their applications have rapidly expanded into the field of Electronic Design Automation (EDA), demonstrating great potential in hardware code generation and verification. In the domain of LLM-based hardware verification, existing research primarily focuses on two directions: testbench generation~\cite{zhang2025llm4dv,ma2024verilogreader,qiu2025correctbench} and assertion generation~\cite{yan2025assertllm,kang2025fveval,wang2025deepassert}. These studies demonstrate the potential of LLMs to understand hardware semantics and verification constraints. AutoBench~\cite{qiu2024autobench} proposed an end-to-end verification framework where LLMs are used to generate all necessary verification artifacts, from test plans to testbenches. Building on this framework, CorrectBench~\cite{qiu2025correctbench} and Pro-V~\cite{zhao2025pro} extended LLM-based testbench generation with additional mechanisms for validating the correctness of the generated testbenches. However, these methods primarily target IP-level designs and are not readily applicable to processor-scale designs. Additionally, ChatFuzz~\cite{rostami2024beyond} fine-tuned GPT-2 for CPU fuzzing  and reported improved fuzzing effectiveness. However, this approach treats the model merely as a seed generator, failing to fully exploit the model's potential for semantic guidance.

\begin{figure}[t]
  \centering
  \small            % ← 整个 figure 内默认用 small，比正文小一号

  \begin{flushleft}
  \footnotesize \textbf{Code:}
  \end{flushleft}

  \begin{lstlisting}[
    language=Verilog,
    basicstyle=\ttfamily\footnotesize,   % ← 代码主体再缩小一档
    keywordstyle=\bfseries\color{blue}, 
    frame=single,
    numbers=left,
    numberstyle=\tiny\color{gray}       % ← 行号更小
  ]
@\textcolor{blue}{\%000010}@ if (mem_resp_valid & traverse & ...) begin
@\textcolor{blue}{\%000002}@   if (3'h0 == r) begin 
@\textcolor{blue}{\%000001}@     data__0 <= pte_ppn[19:0]; 
           // ...
@\textcolor{red}{\%000000}@   if (3'h4 == r) begin 
@\textcolor{red}{\%000000}@     data__4 <= pte_ppn[19:0]; 
@\textcolor{red}{\%000000}@   if (3'h5 == r) begin 
@\textcolor{red}{\%000000}@     data__5 <= pte_ppn[19:0]; 
           // ...
@\textcolor{red}{\%000000}@   if (3'h7 == r) begin 
@\textcolor{red}{\%000000}@     data__7 <= pte_ppn[19:0]; 
          // ...
  \end{lstlisting}

  \begin{flushleft}
  \footnotesize \textbf{LLM reasoning:}
  \end{flushleft}

  % 把解释文字整体缩到 footnotesize
  {\footnotesize
  \par\noindent\hspace{\lstoffset}\fbox{%
    \parbox{\dimexpr\columnwidth-2\fboxsep-2\fboxrule-2\lstoffset\relax}{%
        This code segment resides within the \textcolor{purple}{Page Table Walker (PTW)} and handles cache refills upon receiving a response during page table traversal. The variable r is determined by the cache replacement policy: if all cache entries are valid, the \textcolor{purple}{Pseudo-LRU (PLRU)} policy selects the replacement entry; otherwise, the first invalid entry is chosen. \textcolor{purple}{To cover the target code paths}, we need to trigger \textcolor{purple}{cache refill} where r takes values of 4, 5, 6, or 7. The testcase is: ...
    }%
  }
  }

  \caption{ LLM-guided reasoning for generating testcases targeting uncovered (red) code regions.}
  \label{casestudy}
\end{figure}

\begin{comment}
\begin{figure}[htbp]
\includegraphics[width=0.8\linewidth]{figure/case.pdf} % Replace with the path to your image
    \caption{Hardware Code Comprehension with LLMs .}
    \label{fig:case}
    
\end{figure}
\end{comment}
\subsection{Motivation}
\textbf{Motivation I: Existing hardware fuzzers are semantics-blind, whereas effective hardware verification requires semantic reasoning over DUT behavior.}

\textbf{(1) LLM as a fuzzing generator.}
Traditional coverage-guided fuzzing relies on structural feedback and random mutation, but does not explicitly reason about the control and state conditions required to reach hard-to-trigger hardware behaviors. As shown in Figure~\ref{casestudy}, this limits its ability to systematically construct testcases that satisfy branch conditions such as \texttt{3'h5 == r}. In particular, mutation-based strategies cannot directly identify the state-preparation steps needed to drive execution toward such conditions. By analyzing the code, LLMs can infer that this module manages the PTE cache using the PLRU replacement policy, and can further reason that: (1) the condition \texttt{3'h5 == r} requires both cache saturation and the PLRU state to point to index 5; (2) constructing this state requires specific virtual-address access sequences to ``warm'' and ``cool'' cache entries, thereby steering the internal PLRU state toward the desired index. Leveraging this semantic understanding, LLMs generate targeted testcases that effectively explore uncovered paths, overcoming the inherent limitations of traditional fuzzing approaches.

\textbf{(2) LLM as a fuzzing mutator.}
  Traditional fuzzing mutators generate testcases by applying rule-based transformations to inputs sampled from the seed corpus.
  This approach suffers from fundamental limitations: due to its inability to comprehend the semantics of testcases, it is restricted to simple mutations, such as operand swapping and instruction reordering. Furthermore, while attempting complex transformations, the mutators often introduce syntactic errors. Consequently, such mutators cannot integrate and reorganize testcases with different semantics, making it difficult to trigger complex defect scenarios that require the combination of multiple semantic features. In contrast, LLM-based mutators can combine information from semantically different testcases at a higher level of abstraction and typically preserve syntactic validity and semantic coherence better than rule-based mutation, making them a more flexible mechanism for constructing bug-oriented candidates.

\textbf{Motivation II: Historical bug information provides useful priors for discovering new bugs. The demonstrated effectiveness of leveraging such information in software fuzzing motivates its incorporation into hardware fuzzing to improve bug discovery efficiency.}

Leveraging historical bug data, such as failure traces and patches, has proven highly effective in guiding fuzzers toward new bug discovery within software fuzzing ~\cite{holler2012fuzzing,deng2023large,park2020fuzzing}. Inspired by this success, we observe a similar pattern in hardware design: modules that have previously exhibited bugs are often prone to recurring or related faults. Consequently, utilizing historical bug-triggering artifacts in hardware fuzzing allows us to focus test generation on these vulnerable areas, thereby enhancing overall fuzzing efficiency.

\section{Method}

\begin{figure*}[htbp!] % 单栏浮动体，优先顶端
  \centering
  \includegraphics[width=\linewidth]{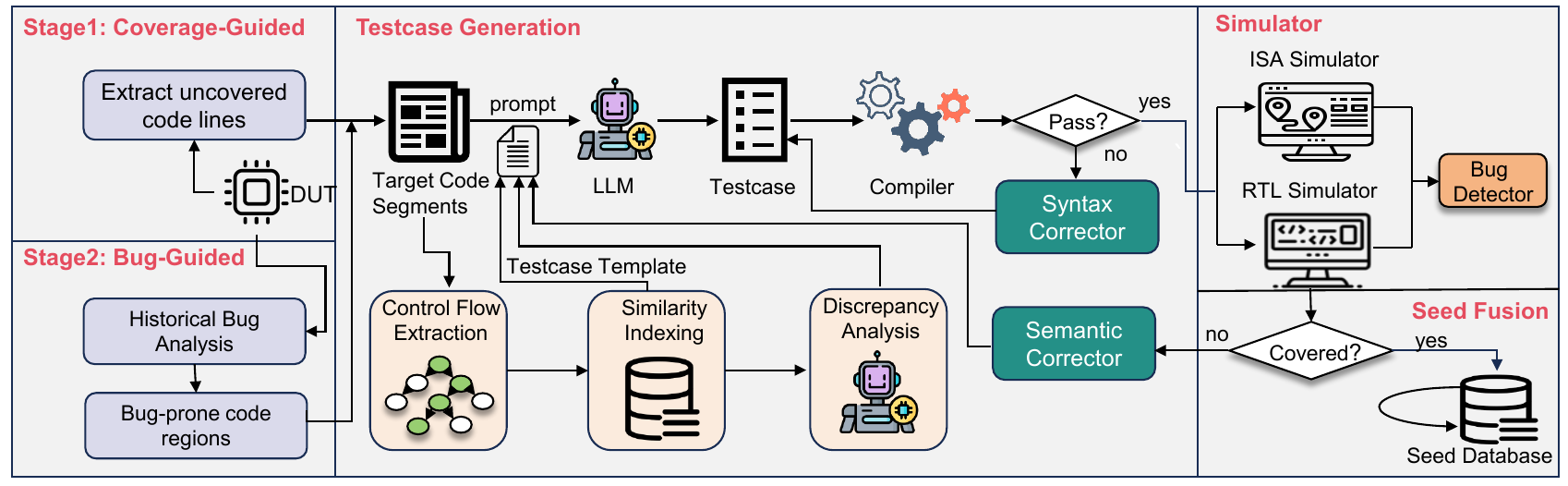} 
  \caption{Overview.}
  \label{fig:overview}
\end{figure*}

\subsection{Overview}
ChipFuzzer adopts a dual-stage workflow, as illustrated in Figure~\ref{fig:overview}. It begins with a coverage-guided stage and then transitions to a bug-guided stage once the coverage improvement of Stage~I becomes marginal, as defined in Section~5.1.

In Stage~I, ChipFuzzer performs coverage-guided fuzzing.  It first selects target code segments from uncovered code lines in the DUT. For each selected target code segment, ChipFuzzer retrieves from the seed database the testcase template whose exercised control-flow path is most similar to the target path, and then performs discrepancy analysis between the target path and the path exercised by the retrieved testcase. The target code segment, the retrieved testcase template, and the discrepancy-analysis results are incorporated into a prompt for the LLM to generate a candidate testcase. The generated testcase is first compiled by the RISC-V toolchain. Compilation failures are repaired by the syntax corrector, whereas executable testcases that still fail to cover the target code segment are refined by the semantic corrector. The corrected testcase is then executed on both RTL and ISA simulators, and a detector module checks whether the target code segment has been covered and whether potential design bugs have been exposed. Testcases that successfully cover the target code segment are stored in the coverage-guided seed database.

In Stage~II, ChipFuzzer performs bug-guided fuzzing. Instead of selecting targets from uncovered code regions, it identifies bug-prone code regions using three strategies derived from historical bug data and then selects target code segments from these regions. Stage~II reuses the same testcase generation, correction, simulation, and detection pipeline as Stage~I. Testcases that successfully cover the selected target code segments are stored in the bug-guided seed database. Finally, ChipFuzzer applies a semantic-aware seed fusion strategy to combine the seeds collected from the two stages.

\subsection{Coverage-Guided Fuzzing}
Coverage-guided fuzzing improves coverage through an iterative process driven by coverage feedback. At the beginning of Stage I, ChipFuzzer uses a small set of simple testcases to bootstrap the seed database. These testcases are first executed on the DUT to obtain their exercised control-flow paths and initial coverage reports, and each testcase is then stored together with its control-flow path as an initial seed entry. Coverage reports collected from RTL simulation are used to identify uncovered code lines in the DUT.  ChipFuzzer then selects target code segments from the uncovered lines, prioritizing semantically related contiguous segments whenever possible; when no such segments are available, it falls back to random selection among the remaining uncovered lines. These target code segments are then passed to the testcase generation pipeline (Section~4.2.1) to produce targeted testcases. The resulting testcases are executed and validated in subsequent iterations, forming a closed loop of coverage analysis, target selection, testcase generation, and simulation validation that progressively improves coverage.

\subsubsection{Testcase Generation Pipeline.}
The testcase generation pipeline takes a selected target code segment as input. It first uses Pyverilog~\cite{pyverilog} to extract the corresponding control-flow path, and then applies the path similarity indexing method (Section~4.2.2) to retrieve the testcase with the highest control-flow similarity from the seed database. The retrieved testcase serves as the template for prompt construction.
ChipFuzzer then performs discrepancy analysis (Section~4.2.3) to compare the target path with the path exercised by the retrieved testcase, thereby identifying the missing conditions or behaviors required to reach the target code segment. The target code segment, the retrieved testcase  template, and the discrepancy analysis results are jointly incorporated into the prompt for the large language model. For Chisel-based CPU designs, the corresponding Chisel source code is also incorporated into the prompt to provide additional semantic context. The final prompt fed to the LLM is shown in Figure~\ref{fig:prompt}.

\begin{figure}[t]
  \centering
  \small
  \begin{flushleft}
You are an RTL verification engineer. Generate a compilable RISC-V testcase to cover the uncovered target code below. Please revise the retrieved testcase template according to the discrepancy analysis. 
  \end{flushleft}

  \begin{lstlisting}[
    language=Verilog,
    basicstyle=\ttfamily\footnotesize,
    keywordstyle=\bfseries\color{blue},
    frame=single
  ]
if (mem_resp_valid & traverse & ...) begin
    if (3'h4 == r) begin
        data__4 <= pte_ppn[19:0];
    if (3'h5 == r) begin
        data__5 <= pte_ppn[19:0];
        // ...
    if (3'h7 == r) begin
        data__7 <= pte_ppn[19:0];
  \end{lstlisting}

  \begin{flushleft}
The corresponding Chisel/Scala context code is:

  \end{flushleft}

  \begin{lstlisting}[
    language=Verilog,
    basicstyle=\ttfamily\footnotesize,
    keywordstyle=\bfseries\color{blue},
    frame=single
  ]
when (mem_resp_valid && traverse && can_refill
&& ...) {
      ...
      data(r) := pte.ppn
      plru.access(r)
    }
  \end{lstlisting}

  \begin{flushleft}
The retrieved testcase template is:
  \end{flushleft}

  \begin{lstlisting}[
    basicstyle=\ttfamily\footnotesize,
    frame=single
  ]
mv      a5, a0
...
or      a5, a4, a5
  \end{lstlisting}

  \begin{flushleft}
Discrepancy analysis results:
  \end{flushleft}

\begin{lstlisting}[
  basicstyle=\ttfamily\footnotesize,
  frame=single,
  breaklines=true,
  breakatwhitespace=true,
  columns=fullflexible
]
- first divergence point: shared prefix: if (mem_resp_valid && traverse && ...); first divergence: if (3'h5 == r)
- missing branch/control condition: the target path requires the PTW replacement index r = 5
- required state preparation: write satp, execute sfence.vma, install valid PTEs, load selected virtual-page bases into a0-a3, ... fill PTW cache, ... bias the PLRU state toward victim index 5
- required trigger event: load a new virtual-page address into a4, issue ld/sw through a4 to trigger a PTW miss, ... so that the returned PTE response enters the refill path
- legality requirement: preserve valid RISC-V syntax, legal PTEs, correct privilege mode, and consistent satp / CSR settings
\end{lstlisting}

  \caption{Prompt template for testcase generation.}

  \label{fig:prompt}
\end{figure}

After the LLM generates a candidate testcase, ChipFuzzer first compiles it with the RISC-V toolchain. Compilation failures are repaired by the syntax corrector, whereas a testcase that compiles successfully but still fails to cover the target code segment is revised by the semantic corrector based on discrepancy analysis and execution feedback (Section~4.2.4). The corrected testcase is then executed on both the RTL and the ISA simulators, and its outputs are subjected to differential analysis by a detector module to identify potential design flaws. If the testcase successfully covers the selected target code segment, the testcase together with its exercised control-flow path is stored in the corresponding seed database. Notably, a testcase that fails to cover the selected target code segment may still cover other previously uncovered regions. Although such a testcase is regarded as a failure with respect to the current target, it is still retained in the seed database because it may serve as a useful seed for subsequent testcase generation and exploration.

\subsubsection{Path Similarity Indexing.}
For each selected target code segment, ChipFuzzer uses Pyverilog~\cite{pyverilog} to extract the corresponding target control-flow path and compares it with the control-flow paths stored in the seed database. Each database entry stores a testcase together with the control-flow path exercised by that testcase. The similarity between the target path and a database path is quantified as follows:

\[
S_{\text{similarity}} = \frac{N_{\text{common}}}{N_{\text{target}}}
\]

where \(N_{\text{common}}\) denotes the number of common nodes shared by the two paths, and \(N_{\text{target}}\) denotes the number of nodes in the target path. ChipFuzzer retrieves the testcase with the highest \(S_{\text{similarity}}\) score as the template for subsequent discrepancy analysis and testcase generation. Among testcases with identical \(S_{\text{similarity}}\) scores, preference is given to the testcase with lower achieved coverage, since such a testcase typically contains fewer incidental behaviors unrelated to the target path and thus provides a cleaner template for subsequent path-specific generation.

\subsubsection{Discrepancy Analysis.}
Path similarity indexing retrieves the seed whose exercised path is most similar to the target path. However, this retrieved seed still does not satisfy all the conditions required to reach the target path. ChipFuzzer uses the retrieved testcase as a reference template and performs discrepancy analysis between the target path and the path exercised by the retrieved testcase. This analysis is guided by five explicit rules, enabling the LLM to identify the missing execution conditions in a structured manner.

Specifically, discrepancy analysis is guided by five rules, as follows. The \textbf{path-alignment rule} aligns the target path with the retrieved execution path and identifies  their first divergence point. The \textbf{branch-condition rule} determines the branch predicates or control conditions required to follow the target path from the divergence point onward.
 The \textbf{state-preparation rule} infers the register values, CSR settings, memory contents, privilege states, or other architectural states that must be prepared before the target path can be reached. The \textbf{event-trigger rule} determines whether exceptions, interrupts, handshakes, valid-response signals, or other trigger events are required to activate the target behavior. The \textbf{legality rule} ensures that the inferred modifications remain consistent with ISA semantics and compilation constraints. Following these rules, the LLM produces a structured discrepancy analysis that explains why the retrieved testcase fails to reach the target code segment and what conditions remain unsatisfied. This discrepancy analysis is then incorporated into prompt construction for subsequent testcase generation.

\subsubsection{Syntax and Semantic Correction.}
ChipFuzzer includes two correction mechanisms for generated testcases. The syntax corrector repairs compilation failures by feeding compiler diagnostics back to the LLM. In practice, syntax failures of LLM-generated testcases mainly arise from low-level assembly issues, such as invalid register names, unsupported operand combinations, malformed instruction formats, unresolved labels and branch targets, improper CSR accesses, and privilege-inconsistent instructions. To reduce such errors, ChipFuzzer further provides the LLM with relevant ISA specification information, including legal register names, CSR definitions, operand-format constraints, and privilege-related instruction requirements. After these errors are repaired, the testcase is recompiled and executed again in the next iteration. The semantic corrector is invoked for executable testcases that still fail to cover the target code segment. It starts from the discrepancy analysis used in the initial testcase generation and refines that analysis according to the execution feedback returned by simulation. Guided by the updated discrepancy analysis, the LLM revises the existing testcase rather than regenerating it from scratch. The revised testcase is then recompiled and re-executed in the next iteration.

\subsection{Bug-Guided Fuzzing}
Bug-guided fuzzing improves bug-finding effectiveness by steering testcase generation toward bug-prone code regions. The workflow proceeds as follows. First, ChipFuzzer employs three strategies based on historical bug data to identify bug-prone code regions (Section~4.3.1). Next, target code segments are selected from these regions and passed to the testcase generation pipeline (Section~4.2.1) to generate testcases. Unlike Stage I, which primarily aims to improve coverage, Stage II focuses on targeted testing of  these historically bug-prone regions. During this stage, testcases that successfully cover the selected target code segments are stored in the bug-guided seed database.
 Finally, ChipFuzzer applies the seed fusion strategy (Section~4.3.2) to fuse the coverage-guided and bug-guided seeds, thereby producing semantically richer testcases for exposing potential design flaws in the DUT.

\subsubsection{Bug-Prone Code Regions.}
Prior work in software fuzzing has shown that historical bug information can provide effective priors for improving bug-finding efficiency~\cite{holler2012fuzzing,deng2023large,park2020fuzzing,zhong2022enriching}. Motivated by this observation, ChipFuzzer incorporates historical bug priors into hardware fuzzing. To this end, we construct a corpus of 128 validated bug-related pull requests (PRs) collected from several open-source RISC-V CPU projects through keyword-based retrieval followed by manual inspection. We then use this corpus to derive bug priors for identifying bug-prone code regions in bug-guided fuzzing. Specifically, ChipFuzzer derives bug priors from three complementary sources: historical bug instances, module-level recurrent bug patterns, and signal-level confusion patterns.

\textbf{Historical Bug Instances.}
We analyze the 128 validated bug-related PRs to extract code segments directly associated with historical defects, and treat these segments as explicit bug-prone targets. Based on these targets, the LLM generates testcases that reproduce or approximate the corresponding bug-triggering scenarios. The resulting testcases are further mutated to exercise nearby code regions that are semantically related to the historical fixes.

\textbf{Module-Level Recurrent Bug Patterns.}
In our collected corpus, bug-fixing changes are not uniformly distributed across the design hierarchy. For example, approximately 20\% of the validated bug-related PRs are associated with cache-related logic, while interconnect- and bus-protocol-handling logic also appears repeatedly in the corpus. Based on these observations, ChipFuzzer captures such recurrence patterns as module-level bug priors and assigns higher priority to code segments in modules that appear more frequently in historical bug fixes, such as cache controllers, coherence logic, and protocol-handling components.

\textbf{Signal-Level Confusion Patterns.}
We also observe that many validated bug-related PRs involve signal-level mistakes. Within our corpus, 85 out of 128 cases involve incorrect signal usage, mismatched control signals, or erroneous signal interactions. ChipFuzzer treats these recurring signal-level error patterns as an additional source of bug priors and prioritizes code segments involving critical control signals. Such signals frequently arise in protocol transitions, exception handling, privilege control, and valid/ready handshakes, where design errors may be more likely to surface.

\subsubsection{Seed Fusion.}
Coverage-guided seeds and bug-guided seeds provide different but complementary information for testcase construction. Coverage-guided seeds capture useful execution patterns and state-setup behaviors from prior exploration, whereas bug-guided seeds capture bug-relevant conditions and behaviors associated with historically bug-prone regions. ChipFuzzer therefore applies a semantic-aware seed fusion strategy to combine the structural usefulness of the former with the bug-oriented information of the latter.

For seed fusion, ChipFuzzer first forms two candidate pools. The coverage-guided candidate pool consists of validated seeds collected during the coverage-guided stage, and the bug-guided candidate pool consists of validated seeds collected during the bug-guided stage that cover the current bug-prone region. ChipFuzzer then samples one seed from each candidate pool to form a fusion pair.

Given the sampled seed pair, ChipFuzzer constructs a fusion prompt that presents the two seeds as complementary references for the LLM. The prompt instructs the LLM to preserve useful instruction subsequences and state-setup patterns from the coverage-guided seed, while selectively incorporating bug-related conditions, trigger events, and control-flow-steering patterns from the bug-guided seed. Rather than directly concatenating the two seeds, ChipFuzzer performs fusion at the semantic level by asking the LLM to rewrite them into a single compilable testcase. If the fused testcase fails compilation, it is repaired by the syntax correction pipeline described in Section~4.2.4.

\section{Evaluation}
\subsection{Experiment Setup}

We evaluate ChipFuzzer on three widely used open-source RISC-V processors, summarized in Table~\ref{tab:benchmarks}: RocketCore~\cite{Asanović:EECS-2016-17} and BOOM~\cite{Celio:EECS-2015-167}, both implemented in Chisel~\cite{bachrach2012chisel}, and CVA6~\cite{zaruba2019cost}, implemented in SystemVerilog. These benchmarks cover both in-order and out-of-order microarchitectures and span different design scales, making them suitable for evaluating the generality of ChipFuzzer across diverse CPU implementations. For the evaluation infrastructure, we use PyVerilog for RTL parsing and control-flow extraction, Verilator~\cite{verilator2025} for RTL simulation, and Spike~\cite{spike} as the ISA golden reference model. All experiments are run on Linux servers with Intel Xeon Silver 4314 (2.40 GHz) processors.

To assess the impact of model selection on coverage performance, we instantiate ChipFuzzer with three large language models: GPT-5.3, Gemini 3, and DeepSeek-V3. We set the maximum number of iterations for both syntax correction and semantic correction to five. ChipFuzzer transitions from Stage I to Stage II when the cumulative condition-coverage gain over 100 consecutive validated testcases falls below 0.1\%.  We evaluate performance using three coverage metrics: register-toggle coverage, condition coverage, and multiplexer-toggle coverage. We select TheHuzz, Cascade, and BMCFuzz as the baselines in our experimental comparison. ChatFuzz is not included because its implementation is not publicly available, making it difficult to conduct a fair and reproducible evaluation under the same experimental setting. We further assess bug detection capability on the Encarsia~\cite{bolcskei2025encarsia} CPU fuzzing platform, which provides a high-quality corpus of buggy hardware designs for fair and reproducible evaluation. The Encarsia bug set used in our evaluation is strictly excluded from the historical bug corpus used for bug-prior construction. Specifically, there is no bug-instance overlap between the Encarsia evaluation set and the 128 collected bug-related PRs. 
\begin{table}[t]
\centering
\caption{Benchmarks used in our study. OoO denotes out-of-order execution, and BP denotes branch prediction.}
\small
\setlength{\tabcolsep}{3pt}
\begin{tabular}{@{}lcccccc@{}}
\toprule
\textbf{Processor} & \textbf{Lang.} & \textbf{\# Latches} & \textbf{\# Gates} & \textbf{Pipeline} & \textbf{OoO} & \textbf{BP} \\
\midrule
RocketCore & Chisel & $1.64 \times 10^{5}$ & $9.24 \times 10^{5}$ & 5-stage & No  & Yes \\
BOOM       & Chisel & $1.99 \times 10^{5}$ & $1.26 \times 10^{6}$ & 7-stage & Yes & Yes \\
CVA6       & SV     & $2.48 \times 10^{4}$ & $4.63 \times 10^{5}$ & 6-stage & No  & Yes \\
\bottomrule
\end{tabular}
\label{tab:benchmarks}
\end{table}

\subsection{Coverage Evaluation}

\begin{figure*}[htbp]
\includegraphics[width=\linewidth]{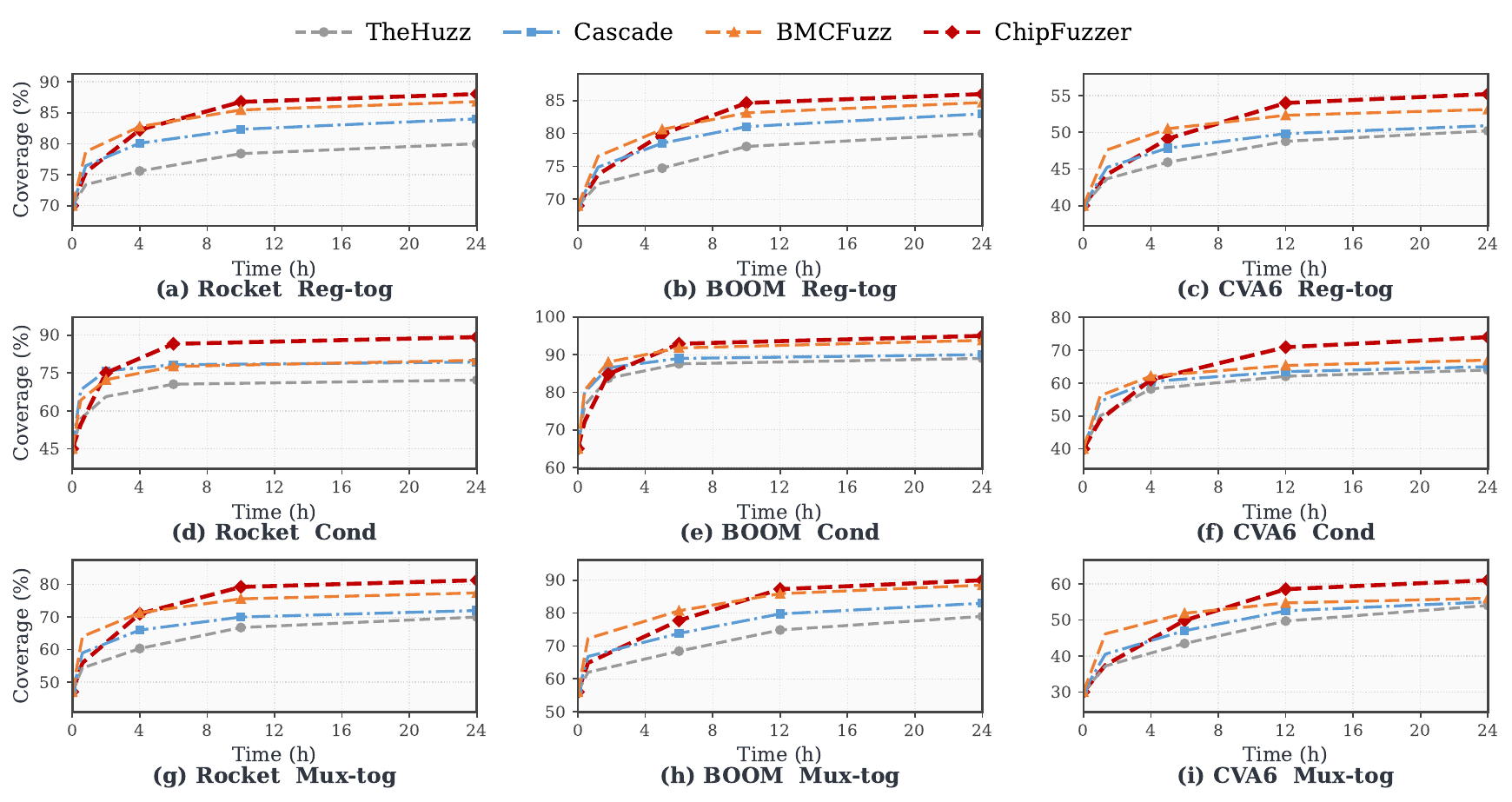} % Replace with the path to your image
    \caption{Coverage growth of ChipFuzzer and baseline methods under register-toggle, condition, and multiplexer-toggle coverage on RocketCore, BOOM, and CVA6.}

    \label{fig:evaluation_baseline}
    
\end{figure*}

We evaluate the coverage effectiveness of ChipFuzzer under a 24-hour fuzzing budget~\cite{bolcskei2025encarsia}. As shown in Figure~\ref{fig:evaluation_baseline}, ChipFuzzer achieves the best coverage performance on RocketCore, BOOM, and CVA6 under all three metrics. For condition coverage, ChipFuzzer improves over the strongest baseline by 9.1, 1.2, and 7.0 percentage points on RocketCore, BOOM, and CVA6, respectively. On BOOM, existing baselines already achieve high condition coverage, so the remaining room for improvement is limited. In contrast, the baseline methods attain substantially lower condition coverage on CVA6, while ChipFuzzer still improves over the strongest baseline by 7.0 percentage points. This result highlights the advantage of semantic testcase generation when reaching complex control conditions is critical for improving coverage. ChipFuzzer also outperforms the baseline methods in register-toggle coverage and multiplexer-toggle coverage. This shows that ChipFuzzer not only covers more control conditions, but also activates more internal signals and data paths.

We also observe from the coverage curves in Figure~\ref{fig:evaluation_baseline} that ChipFuzzer increases coverage more slowly in the early fuzzing stage. Its testcase  generation throughput is constrained by the latency of LLM inference and is therefore lower than that of mutation-based baselines. However, once the coverage gains of the baseline methods start to plateau, ChipFuzzer continues to make steady progress and eventually achieves the best overall coverage. These observations suggest that ChipFuzzer is complementary to existing fuzzing approaches. Mutation-based fuzzing can be used in the early stage to rapidly explore easy-to-reach states, and ChipFuzzer can then be applied in the later stage to further improve coverage and bug-finding effectiveness.

\begin{comment}

\begin{figure*}[htbp]
\includegraphics[width=\linewidth]{figure/coverage-baseline-growth1-abl.pdf} % Replace with the path to your image
    \caption{Coverage evaluation with ablation study.}
    \label{fig:evaluation_ablation}
    
\end{figure*}

\end{comment}

We further assess testcase efficiency by measuring the number of testcases required to reach a target coverage threshold, as summarized in Table~\ref{tab:coverage75}. ChipFuzzer requires substantially fewer testcases than BMCFuzz, Cascade, and TheHuzz on both RocketCore and BOOM. On RocketCore, ChipFuzzer reaches 75\% coverage with only 645 testcases, corresponding to 6.7$\times$, 9.2$\times$, and 26.6$\times$ fewer testcases than BMCFuzz, Cascade, and TheHuzz, respectively. On BOOM, ChipFuzzer reaches 90\% coverage with 734 testcases, whereas the three baselines require 7.6$\times$, 10.5$\times$, and 31.3$\times$ more testcases. These results show that ChipFuzzer achieves substantially higher coverage efficiency than the baseline methods. This advantage stems from its ability to steer testcase generation toward uncovered code regions and the control conditions required to reach them, allowing it to reach the same coverage with far fewer testcases.

\begin{table}[t]
\centering
\caption{Number of testcases required to achieve the condition coverage
(75\% on RocketCore and 90\% on BOOM). Lower values indicate higher
efficiency.}
\small
\begin{tabular}{@{}lccc@{}}
\toprule
\textbf{Processor} & \textbf{Method} & \textbf{Testcase Count}& \textbf{Rel. to ChipFuzzer} \\
\midrule
\multirow{4}{*}{RocketCore (75\%)} 
  & ChipFuzzer & 645      & --  \\
  & BMCFuzz    & 4{,}352  & 6.7$\times$ \\
  & Cascade    & 5{,}946  & 9.2$\times$ \\
  & TheHuzz    & 17{,}126 & 26.6$\times$ \\
\midrule
\multirow{4}{*}{BOOM (90\%)} 
  & ChipFuzzer & 734      & -- \\
  & BMCFuzz    & 5{,}600  & 7.6$\times$ \\
  & Cascade    & 7{,}694  & 10.5$\times$ \\
  & TheHuzz    & 23{,}005 & 31.3$\times$ \\
\bottomrule
\end{tabular}
\label{tab:coverage75}
\end{table}

\begin{figure}[htbp]
    \centering
    \includegraphics[width=\linewidth]{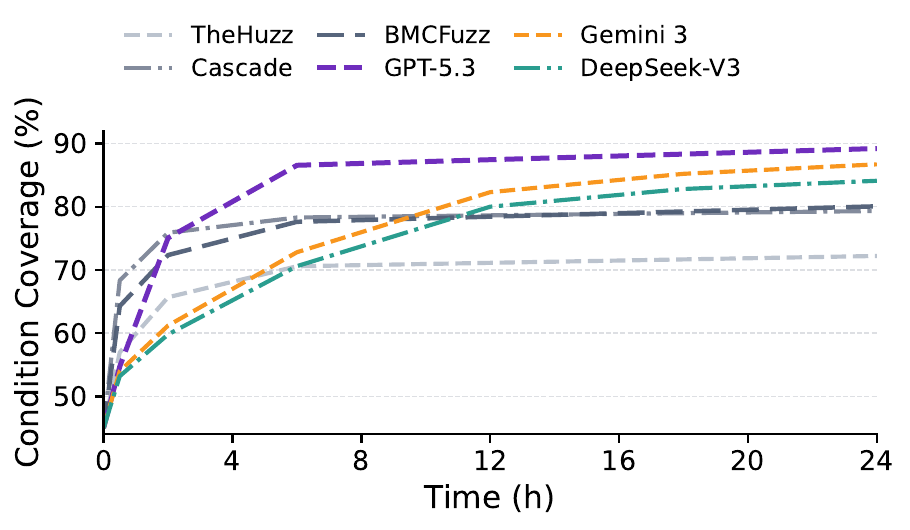}
    \caption{Condition coverage comparison of different foundation models and baseline methods on RocketCore.}
    \label{fig:evaluation_model_rocket}
\end{figure}

We further evaluate the impact of the underlying foundation model by instantiating ChipFuzzer with three foundation models, namely GPT-5.3, Gemini 3, and DeepSeek-V3, while keeping the prompting strategy, fuzzing budget, and execution pipeline unchanged. As shown in Figure~\ref{fig:evaluation_model_rocket}, the choice of foundation model has a clear impact on the final condition coverage achieved by ChipFuzzer on RocketCore. Among the evaluated models, GPT-5.3 achieves the best result with 89.2\% condition coverage, followed by Gemini 3 with 86.7\% and DeepSeek-V3 with 84.1\%. These results suggest that the choice of foundation model affects the final coverage achieved by ChipFuzzer, while all three ChipFuzzer variants still outperform the baseline methods.

\begin{comment}
    
\end{comment}

\subsection{Bug Evaluation}
Evaluating fuzzing tools based on naturally occurring defect rates presents challenges due to variations in hardware designs and versions, which complicate direct comparisons. To ensure a fair assessment, we employ the open-source vulnerability corpus from Encarsia, strictly adhering to their experimental configuration. This benchmark comprises 90 bugs (30 per design) across three CPU architectures. ChipFuzzer achieves an average  bug  detection rate of 61.1\%, outperforming TheHuzz  (36.7\%), Cascade (40\%), BMCFuzz (40\%), as shown in  Figure~\ref{fig:bug}. These results indicate that ChipFuzzer significantly enhances bug-finding effectiveness by integrating historical bug knowledge into semantically guided testcase generation, and further suggest that the methodology of leveraging historical bug information in software fuzzing remains applicable to hardware fuzzing.

\newcommand{\LO}{\textsc{LO}}
\newcommand{\CG}{\textsc{CG}}
\newcommand{\BG}{\textsc{BG}}
\newcommand{\CBG}{\textsc{CBG}}

\subsection{Ablation Study}
In this section, we conduct an ablation study to assess the contributions of individual components in ChipFuzzer. We evaluate the impact on both coverage and bug-finding performance under four experimental configurations:
\begin{enumerate}[leftmargin=*, itemsep=0.25ex, topsep=0.4ex]
  \item \textbf{LLM Only (\LO).} The foundational baseline configuration, where fuzzing is driven by the LLM without additional guidance.
  \item \textbf{Coverage-Guided (\CG).} Fuzzing with only the coverage-guided stage active.
  \item \textbf{Bug-Guided (\BG).} Fuzzing with only the bug-guided stage active.
  \item \textbf{Coverage and Bug-Guided (\CBG).} Fuzzing with both coverage-guided and bug-guided stages active.
\end{enumerate}

\begin{figure}[t]
    \centering
    \includegraphics[width=\linewidth]{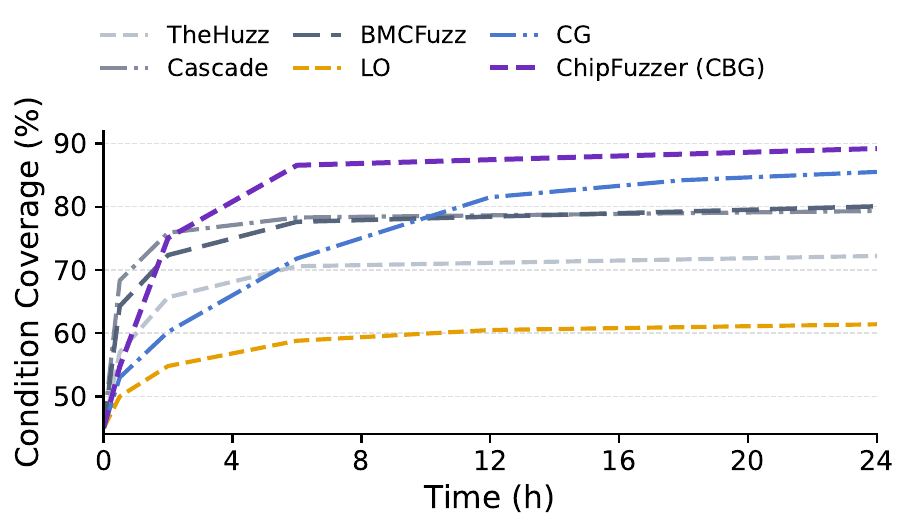}
    \caption{Ablation results on RocketCore under condition coverage.}
    \label{fig:ablation_rocket_cond}
\end{figure}

\textbf{Coverage Evaluation.}
We evaluate the ablation results on RocketCore under condition coverage, with the results shown in Figure~\ref{fig:ablation_rocket_cond}. The \textbf{BG} configuration is omitted in this part, as its objective is bug discovery rather than coverage maximization. The \textbf{LO} configuration, which relies solely on unguided LLM-based testcase generation, reaches 61.4\% condition coverage. This result remains below TheHuzz (72.21\%), Cascade (79.32\%), and BMCFuzz (80.06\%), indicating that unguided generation alone is insufficient for covering the more complex control conditions in the design. Enabling \textbf{CG} raises condition coverage to 88.1\%, surpassing all three baselines and validating the effectiveness of the control-flow-aware guidance and discrepancy analysis adopted in ChipFuzzer. These mechanisms steer testcase generation toward uncovered code regions and the control conditions required to reach them, rather than relying on unguided exploration. Adding bug guidance on top of \textbf{CG} further improves the final coverage to 89.2\%. Although the additional gain is smaller, it indicates that bug guidance provides a complementary benefit once the main coverage-oriented mechanisms are already in place.

\begin{figure}[htbp]
\includegraphics[width=\linewidth]{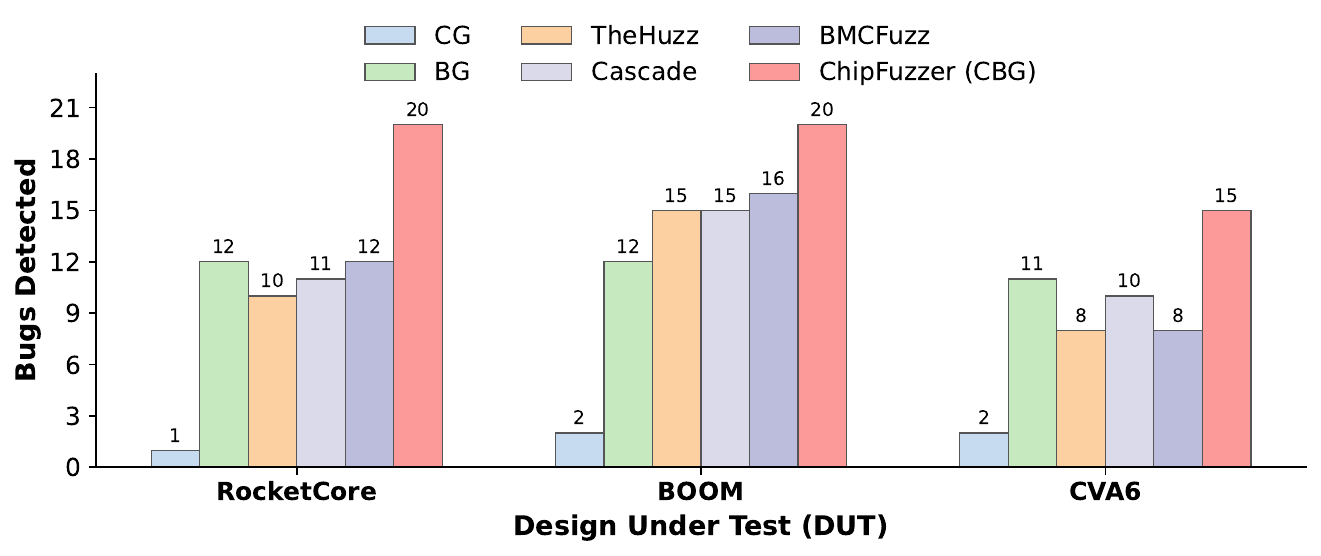} % Replace with the path to your image
    \caption{Bugs detected by baseline fuzzers, ChipFuzzer (CBG), and its CG/BG ablations on three different
    RISC-V designs.}
    \label{fig:bug}
    
\end{figure}

\textbf{Bug Evaluation.}
We evaluate the bug-finding capability of all four configurations, with the results shown in Figure~\ref{fig:bug}. The \textbf{LO} configuration fails to detect any bugs. Enabling only \textbf{CG} yields a bug detection rate of 5.6\%, indicating that coverage improvement alone contributes only limited bug-finding capability in this setting. In contrast, \textbf{BG} raises the detection rate to 38.9\%, confirming that bug-guided knowledge is the dominant factor for directing execution toward bug-prone behaviors. The complete \textbf{CBG} configuration further improves the bug detection rate to 61.1\%, demonstrating that coverage-guided exploration and bug-guided generation provide the strongest bug-finding performance when used together.

\section{Discussion}

ChipFuzzer can serve as an effective complement to traditional hardware fuzzers rather than a direct replacement. Mutation-based fuzzers are efficient at rapidly exploring easy-to-reach states, whereas ChipFuzzer is better suited for generating semantically targeted testcases for hard-to-reach hardware behaviors. This complementarity is also reflected in the coverage trends: although ChipFuzzer progresses more slowly in the early stage due to LLM inference overhead, it continues to improve coverage after conventional fuzzers begin to plateau.

At the same time, several limitations remain. ChipFuzzer depends on the reasoning quality of the underlying LLM and may still hallucinate hardware semantics, especially for custom instructions or insufficiently documented behaviors. In addition, LLM-based generation introduces nontrivial overhead, which limits testcase throughput in the early stage. Finally, our current evaluation focuses on open-source RISC-V CPU designs, and its effectiveness on broader RTL designs requires further investigation. Future work includes combining LLM-guided generation with formal techniques, designing hybrid workflows with fast mutation-based fuzzers, and studying the transferability of bug priors across different designs.

\section{Conclusion}
In this work, we present ChipFuzzer, an LLM-driven hardware fuzzing framework for semantically guided testcase generation. ChipFuzzer adopts a dual-stage workflow comprising a Coverage-Guided stage and a Bug-Guided stage. The Coverage-Guided stage improves exploration by using control-flow similarity and discrepancy analysis to guide testcase generation toward uncovered code regions, while the Bug-Guided stage leverages historical bug data to identify bug-prone regions and prioritize testcase generation toward them. Experimental results on three open-source CPU designs and the Encarsia benchmark show that ChipFuzzer improves coverage and bug detection over strong hardware-fuzzing baselines. These findings demonstrate the promise of large language models for enhancing hardware verification through semantic reasoning and targeted fuzzing.

\bibliographystyle{ACM-Reference-Format}
\bibliography{ref}

@inproceedings{bolcskei2025encarsia,
  title={Encarsia: Evaluating cpu fuzzers via automatic bug injection},
  author={B{\"o}lcskei, Matej and Solt, Flavien and Ceesay-Seitz, Katharina and Razavi, Kaveh},
  booktitle={34th USENIX Security},
  year={2025}
}

@inproceedings{hur2021difuzzrtl,
  title={Difuzzrtl: Differential fuzz testing to find cpu bugs},
  author={Hur, Jaewon and Song, Suhwan and Kwon, Dongup and Baek, Eunjin and Kim, Jangwoo and Lee, Byoungyoung},
  booktitle={2021 IEEE Symposium on Security and Privacy (SP)},
  pages={1286--1303},
  year={2021},
  organization={IEEE}
}

@inproceedings{solt2024cascade,
  title={Cascade:$\{$CPU$\}$ fuzzing via intricate program generation},
  author={Solt, Flavien and Ceesay-Seitz, Katharina and Razavi, Kaveh},
  booktitle={33rd USENIX Security Symposium (USENIX Security 24)},
  pages={5341--5358},
  year={2024}
}

@inproceedings{rostami2024beyond,
  title={Beyond random inputs: A novel ML-based hardware fuzzing},
  author={Rostami, Mohamadreza and Chilese, Marco and Zeitouni, Shaza and Kande, Rahul and Rajendran, Jeyavijayan and Sadeghi, Ahmad-Reza},
  booktitle={2024 Design, Automation \& Test in Europe Conference \& Exhibition (DATE)},
  pages={1--6},
  year={2024},
  organization={IEEE}
}

@inproceedings{kande2022thehuzz,
  title={$\{$TheHuzz$\}$: Instruction fuzzing of processors using $\{$Golden-Reference$\}$ models for finding $\{$Software-Exploitable$\}$ vulnerabilities},
  author={Kande, Rahul and Crump, Addison and Persyn, Garrett and Jauernig, Patrick and Sadeghi, Ahmad-Reza and Tyagi, Aakash and Rajendran, Jeyavijayan},
  booktitle={31st USENIX Security Symposium (USENIX Security 22)},
  pages={3219--3236},
  year={2022}
}

@inproceedings{pyverilog,
  title={Pyverilog: A python-based hardware design processing toolkit for verilog hdl},
  author={Takamaeda-Yamazaki, Shinya},
  booktitle={Applied Reconfigurable Computing: 11th International Symposium, ARC 2015, Bochum, Germany, April 13-17, 2015, Proceedings 11},
  pages={451--460},
  year={2015},
  organization={Springer}
}

@techreport{Asanović:EECS-2016-17,
    Author= {Asanović, Krste and Avizienis, Rimas and Bachrach, Jonathan and Beamer, Scott and Biancolin, David and Celio, Christopher and Cook, Henry and Dabbelt, Daniel and Hauser, John and Izraelevitz, Adam and Karandikar, Sagar and Keller, Ben and Kim, Donggyu and Koenig, John and Lee, Yunsup and Love, Eric and Maas, Martin and Magyar, Albert and Mao, Howard and Moreto, Miquel and Ou, Albert and Patterson, David A. and Richards, Brian and Schmidt, Colin and Twigg, Stephen and Vo, Huy and Waterman, Andrew},
    Title= {The Rocket Chip Generator},
    Year= {2016},
    Month= {Apr},
    Url= {http://www2.eecs.berkeley.edu/Pubs/TechRpts/2016/EECS-2016-17.html},
    Number= {UCB/EECS-2016-17},
}

@techreport{Celio:EECS-2015-167,
    Author= {Celio, Christopher and Patterson, David A. and Asanović, Krste},
    Title= {The Berkeley Out-of-Order Machine (BOOM): An Industry-Competitive, Synthesizable, Parameterized RISC-V Processor},
    Year= {2015},
    Month= {Jun},
    Url= {http://www2.eecs.berkeley.edu/Pubs/TechRpts/2015/EECS-2015-167.html},
    Number= {UCB/EECS-2015-167},
}

@misc{verilator2025,
  author    = {Wilson Snyder},
  title     = {Verilator},
  year      = {2025},
  url       = {https://github.com/verilator/verilator.git},
  note      = {Accessed: 2025-11-14}
}

@misc{spike,
  author    = {Riscv-Software-Src.},
  title     = {Riscv-isa-sim: Spike, a risc-v isa simulator.},
  year      = {2025},
  url       = {https:
 //github.com/riscv-software-src/riscv-isa-sim.},
  note      = {Accessed: 2025-11-14}
}

@article{farahmandi2020system,
  title={System-on-chip security},
  author={Farahmandi, Farimah and Huang, Yuanwen and Mishra, Prabhat},
  journal={Cham, Switzerland: Springer},
  year={2020},
  publisher={Springer}
}

@inproceedings{qiu2024autobench,
  title={Autobench: Automatic testbench generation and evaluation using llms for hdl design},
  author={Qiu, Ruidi and Zhang, Grace Li and Drechsler, Rolf and Schlichtmann, Ulf and Li, Bing},
  booktitle={Proceedings of the 2024 ACM/IEEE International Symposium on Machine Learning for CAD},
  pages={1--10},
  year={2024}
}

@inproceedings{qiu2025correctbench,
  title={Correctbench: Automatic testbench generation with functional self-correction using llms for hdl design},
  author={Qiu, Ruidi and Zhang, Grace Li and Drechsler, Rolf and Schlichtmann, Ulf and Li, Bing},
  booktitle={2025 Design, Automation \& Test in Europe Conference (DATE)},
  pages={1--7},
  year={2025},
  organization={IEEE}
}

@article{zhao2025pro,
  title={PRO-V: An Efficient Program Generation Multi-Agent System for Automatic RTL Verification},
  author={Zhao, Yujie and Wu, Zhijing and Zhang, Hejia and Yu, Zhongming and Ni, Wentao and Ho, Chia-Tung and Ren, Haoxing and Zhao, Jishen},
  journal={arXiv preprint arXiv:2506.12200},
  year={2025}
}

@inproceedings{chen2023hypfuzz,
  title={$\{$HyPFuzz$\}$:$\{$Formal-Assisted$\}$ processor fuzzing},
  author={Chen, Chen and Kande, Rahul and Nguyen, Nathan and Andersen, Flemming and Tyagi, Aakash and Sadeghi, Ahmad-Reza and Rajendran, Jeyavijayan},
  booktitle={32nd USENIX Security Symposium (USENIX Security 23)},
  pages={1361--1378},
  year={2023}
}

@inproceedings{chen2023psofuzz,
  title={PSOFuzz: Fuzzing processors with particle swarm optimization},
  author={Chen, Chen and Gohil, Vasudev and Kande, Rahul and Sadeghi, Ahmad-Reza and Rajendran, Jeyavijayan},
  booktitle={2023 IEEE/ACM International Conference on Computer Aided Design (ICCAD)},
  pages={1--9},
  year={2023},
  organization={IEEE}
}

@inproceedings{gohil2024mabfuzz,
  title={MABFuzz: Multi-armed bandit algorithms for fuzzing processors},
  author={Gohil, Vasudev and Kande, Rahul and Chen, Chen and Sadeghi, Ahmad-Reza and Rajendran, Jeyavijayan},
  booktitle={2024 Design, Automation \& Test in Europe Conference \& Exhibition (DATE)},
  pages={1--6},
  year={2024},
  organization={IEEE}
}

@inproceedings{holler2012fuzzing,
  title={Fuzzing with code fragments},
  author={Holler, Christian and Herzig, Kim and Zeller, Andreas},
  booktitle={21st USENIX Security Symposium (USENIX Security 12)},
  pages={445--458},
  year={2012}
}

@inproceedings{park2020fuzzing,
  title={Fuzzing javascript engines with aspect-preserving mutation},
  author={Park, Soyeon and Xu, Wen and Yun, Insu and Jang, Daehee and Kim, Taesoo},
  booktitle={2020 IEEE Symposium on Security and Privacy (SP)},
  pages={1629--1642},
  year={2020},
  organization={IEEE}
}

@article{zaruba2019cost,
  title={The cost of application-class processing: Energy and performance analysis of a Linux-ready 1.7-GHz 64-bit RISC-V core in 22-nm FDSOI technology},
  author={Zaruba, Florian and Benini, Luca},
  journal={IEEE Transactions on Very Large Scale Integration (VLSI) Systems},
  volume={27},
  number={11},
  pages={2629--2640},
  year={2019},
  publisher={IEEE}
}

@inproceedings{xu2024pathfuzz,
  title={Pathfuzz: Broadening fuzzing horizons with footprint memory for CPUs},
  author={Xu, Yinan and Wang, Sa and Tang, Dan and Sun, Ninghui and Bao, Yungang},
  booktitle={Proceedings of the 61st ACM/IEEE Design Automation Conference},
  pages={1--6},
  year={2024}
}

@inproceedings{laeufer2018rfuzz,
  title={RFUZZ: Coverage-directed fuzz testing of RTL on FPGAs},
  author={Laeufer, Kevin and Koenig, Jack and Kim, Donggyu and Bachrach, Jonathan and Sen, Koushik},
  booktitle={2018 IEEE/ACM International Conference on Computer-Aided Design (ICCAD)},
  pages={1--8},
  year={2018},
  organization={IEEE}
}

@inproceedings{shen2025bmcfuzz,
  title={BMCFuzz: Hybrid Verification of Processors by Synergistic Integration of Bound Model Checking and Fuzzing},
  author={Shen, Shidong and Liu, Jinyu and Feng, Weizhi and Song, Fu and Wu, Zhilin},
  booktitle={2025 IEEE/ACM International Conference On Computer Aided Design (ICCAD)},
  pages={1--9},
  year={2025},
  organization={IEEE}
}

@inproceedings{bachrach2012chisel,
  title={Chisel: constructing hardware in a scala embedded language},
  author={Bachrach, Jonathan and Vo, Huy and Richards, Brian and Lee, Yunsup and Waterman, Andrew and Avi{\v{z}}ienis, Rimas and Wawrzynek, John and Asanovi{\'c}, Krste},
  booktitle={Proceedings of the 49th annual design automation conference},
  pages={1216--1225},
  year={2012}
}

@article{deng2023large,
  title={Large language models are edge-case fuzzers: Testing deep learning libraries via fuzzgpt},
  author={Deng, Yinlin and Xia, Chunqiu Steven and Yang, Chenyuan and Zhang, Shizhuo Dylan and Yang, Shujing and Zhang, Lingming},
  journal={arXiv preprint arXiv:2304.02014},
  year={2023}
}

@inproceedings{zhong2022enriching,
  title={Enriching compiler testing with real program from bug report},
  author={Zhong, Hao},
  booktitle={Proceedings of the 37th IEEE/ACM International conference on automated software engineering},
  pages={1--12},
  year={2022}
}

@inproceedings{cyrluk1994effective,
  title={Effective theorem proving for hardware verification},
  author={Cyrluk, David and Rajan, Sreeranga and Shankar, Natarajan and Srivas, Mandayam K},
  booktitle={International Conference on Theorem Provers in Circuit Design},
  pages={203--222},
  year={1994},
  organization={Springer}
}

@book{clarke2018handbook,
  title={Handbook of model checking},
  author={Clarke, Edmund M and Henzinger, Thomas A and Veith, Helmut and Bloem, Roderick and others},
  volume={10},
  year={2018},
  publisher={Springer}
}

@article{naveh2007constraint,
  title={Constraint-based random stimuli generation for hardware verification},
  author={Naveh, Yehuda and Rimon, Michal and Jaeger, Itai and Katz, Yoav and Vinov, Michael and s Marcu, Eitan and Shurek, Gil},
  journal={AI magazine},
  volume={28},
  number={3},
  pages={13--13},
  year={2007}
}

@article{hu2021hardware,
  title={Hardware information flow tracking},
  author={Hu, Wei and Ardeshiricham, Armaiti and Kastner, Ryan},
  journal={ACM Computing Surveys (CSUR)},
  volume={54},
  number={4},
  pages={1--39},
  year={2021},
  publisher={ACM New York, NY, USA}
}

@inproceedings{trippel2022fuzzing,
  title={Fuzzing hardware like software},
  author={Trippel, Timothy and Shin, Kang G and Chernyakhovsky, Alex and Kelly, Garret and Rizzo, Dominic and Hicks, Matthew},
  booktitle={31st USENIX Security Symposium (USENIX Security 22)},
  pages={3237--3254},
  year={2022}
}

@article{saravanan2024emergence,
  title={The emergence of hardware fuzzing: A critical review of its significance},
  author={Saravanan, Raghul and Dinakarrao, Sai Manoj Pudukotai},
  journal={arXiv preprint arXiv:2403.12812},
  year={2024}
}

@inproceedings{yan2025assertllm,
  title={Assertllm: Generating hardware verification assertions from design specifications via multi-llms},
  author={Yan, Zhiyuan and Fang, Wenji and Li, Mengming and Li, Min and Liu, Shang and Xie, Zhiyao and Zhang, Hongce},
  booktitle={Proceedings of the 30th Asia and South Pacific Design Automation Conference},
  pages={614--621},
  year={2025}
}

@inproceedings{zhang2025llm4dv,
  title={Llm4dv: Using large language models for hardware test stimuli generation},
  author={Zhang, Zixi and Szekely, Balint and Gimenes, Pedro and Chadwick, Greg and McNally, Hugo and Cheng, Jianyi and Mullins, Robert and Zhao, Yiren},
  booktitle={2025 IEEE 33rd Annual International Symposium on Field-Programmable Custom Computing Machines (FCCM)},
  pages={133--137},
  year={2025},
  organization={IEEE}
}

@inproceedings{ma2024verilogreader,
  title={Verilogreader: Llm-aided hardware test generation},
  author={Ma, Ruiyang and Yang, Yuxin and Liu, Ziqian and Zhang, Jiaxi and Li, Min and Huang, Junhua and Luo, Guojie},
  booktitle={2024 IEEE LLM Aided Design Workshop (LAD)},
  pages={1--5},
  year={2024},
  organization={IEEE}
}

@inproceedings{kang2025fveval,
  title={Fveval: Understanding language model capabilities in formal verification of digital hardware},
  author={Kang, Minwoo and Liu, Mingjie and Hamad, Ghaith Bany and Suhaib, Syed M and Ren, Haoxing},
  booktitle={2025 Design, Automation \& Test in Europe Conference (DATE)},
  pages={1--6},
  year={2025},
  organization={IEEE}
}

@article{wang2025deepassert,
  title={DeepAssert: An LLM-Aided Verification Framework with Fine-Grained Assertion Generation for Modules with Extracted Module Specifications},
  author={Wang, Yonghao and Zhou, Jiaxin and Lyu, Hongqin and Chao, Zhiteng and Wang, Tiancheng and Li, Huawei},
  journal={arXiv preprint arXiv:2509.14668},
  year={2025}
}

\end{document}